\documentclass[english,11pt]{amsart}
\usepackage{amsfonts,amsmath,mathrsfs,amssymb}
\usepackage{graphicx}
\usepackage[T1]{fontenc}
\usepackage{latexsym}
\usepackage{varwidth}
\usepackage{a4wide}
\usepackage{hyperref}
\usepackage{caption}
\usepackage{subcaption}
\usepackage{url}
\usepackage{breakurl}

\begin{document}
\baselineskip 6mm
\date{}
\title[The recent dynamical history of comet 67P/Churyumov-Gerasimenko]{The recent dynamical history of comet 67P/Churyumov-Gerasimenko}
\author{Lucie Maquet$^{1}$}

\keywords{chaos, celestial mechanics, comet: individual: 67P/Churyumov-Gerasimenko}

\begin{abstract}
This paper presents the past evolution of the orbital elements of comet 67P/Chu\-ryumov-Gerasimenko, target of the Rosetta spacecraft. The gravitational orbit of the comet is affected by the sublimation of ice from the nucleus that triggers non-gravitational forces. The comet also experienced several close encounters with Jupiter especially in 1959 and 1923 (less than 1 AU). These perturbations cause the chaoticity of the comet orbit at short time scale. The goal of this paper is to have a precise idea of the comet recent dynamical history. This is done in studying the mean trends of the comet orbital elements and also in characterizing its chaotic motion with the Fast Lyapunov indicator. To compute these mean trends, two sets (considering or not the non-gravitational forces) of 1000 clone orbits of the comet were considered. This paper shows that the last close encounter with Jupiter on February $4^{\text{th}}$, 1959 drastically modified the orbital elements of the comet (especially the perihelion distance from more than 2.7 AU prior the encounter to 1.3 AU after the encounter). The motion of the comet is also shown to be chaotic before the close encounter with Jupiter on October $2^{\text{nd}}$, 1923. The mean trends of the orbital elements of the comet are also presented in this paper (for two time scales : 275 years in the past and 10,000 years in the past). 
\end{abstract}

\maketitle

\vskip 5mm
\begin{tiny}
\begin{enumerate}
\item {ESA/ESAC, PO Box 78, 28691 Villanueva de la Ca\~nada, Spain}
\end{enumerate}
\end{tiny}

\tableofcontents

\section{Introduction}
The Rosetta spaceprobe is currently orbiting around the nucleus of the comet 67P/Churyumov-Gerasimenko (67P/C-G in the following text). One of the major goal of its mission is to study the comet during its active phase around the Sun. The orbit of the comet is affected by the sublimation of the ice off the nucleus. When the comet approaches the Sun, this outgassing triggers a non-gravitational force (NGF) that significantly modifies its orbit. Moreover, during its past orbits, comet 67P/C-G had several close encounters with planet Jupiter. Because of these perturbations, the orbit of 67P/C-G is chaotic. This chaoticity was already revealed by \cite{kro2003} by integrating 20 orbits of random clones of the comet and by using different non-gravitational models and \cite{gro2007} by integrating 18 clones with nine different values of the initial mean anomaly and considering the NGF or not.\\

Contrary to these previous studies, the goal of this paper is to show the mean past evolution (taking into account 1000 clones with NGF and another 1000 clones without NGF) of different orbital elements to better understand to better understand what is actually observed by Rosetta. Indeed, providing the mean trends of the orbital elements is necessary for the different instrasec descriptions of the comet such as its formation, its structure and its erosion. In this work, the chaoticity of the comet motion is also highlighted by computing the Fast Lyapunov Indicato rusing the strategy explained in \cite{fou2002} (for a review of this indicator, see \cite{fou2002} and references therein).\\

The paper is organised as follows : \\
Section \ref{model} presents the equation of motion used to integrate the comet motion and how the initial conditions are choosen. Section \ref{results} shows the results of the integrations and how chaotic the motion of the comet is by computing the Fast Lyapunov Indicator. Conclusions are given in section \ref{conclu}.

\section{Modelisation}\label{model}
\subsection{Equation of motion}
\begin{figure}
\begin{center}
\includegraphics[width=0.5\textwidth]{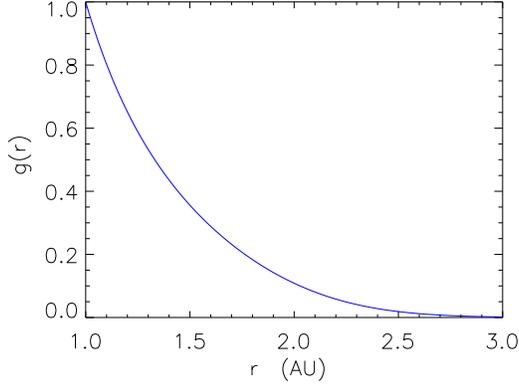}
\caption{$g(r)$ as a function of the heliocentric distance $r$. This figure is generated using the equation \ref{gder}}\label{g_r}
\end{center}
\end{figure}

The dynamical evolution of 67P/C-G is investigated by numerically integrating a two bodies problem (sun and comet) perturbed by the eight planets of the Solar System. The positions of the planets are given by the planetary theory INPOP 13 (\cite{fie2014}) provided by the \textit{Institut de M\'ecanique C\'eleste et de Calcul des \'Eph\'em\'erides}. A solution of this planetary theory covering the time needed by this study (50 000 years) was specially provided for this work by IMCCE. The non-gravitational force due to the cometary outgassing is introduced in the equation of motion as the one of \cite{mar1973}. 
The non-gravitational perturbing acceleration is given by its radial, transverse and normal components, in the equatorial heliocentric frame as :
\begin{equation}
\mathbf{A}_{NG}=A_1 g(r)\mathbf{e}_R+A_2 g(r)\mathbf{e}_T+A_3 g(r)\mathbf{e}_N
\end{equation}
where
\begin{equation}
g(r)=0.111262  \left(\frac{r}{2.808}\right)^{-2.15} \left(1+\left(\frac{r}{2.808}\right)^{5.093}\right)^{-4.6142}\label{gder}
\end{equation}
and $A_1, A_2, A_3$ are constants obtained by fitting the astrometrical positions of the considered comet together with the orbital elements. The vector $\mathbf{e}_R$, $\mathbf{e}_T$ and $\mathbf{e}_N$ are respectively the radial, transverse and normal direction vectors. The dimensionless function $g(r)$ represents the variation in the sublimation rate as a function of the heliocentric distance of the comet $r$. According to figure \ref{g_r} generated from equation \ref{gder}, we note that the activity begins to be significative toward the sun around 2.7 au wich is the snow line of the Solar System (\cite{abe2000,mor2000}). According to \cite{tub2015}, the activity of 67P/C-G began before the snow line but was very weak and not enough strong to affect the dynamic and be detected by orbital computation.\\

The acceleration due to relativistic effects is also introduced because of the proximity of the comet to the sun during the perihelion passage. The relativistic effects are comparable in term of magnitude to the non-gravitational forces (\cite{maq2012}). Moreover, the non-gravitational parameters given in table \ref{tab:ci} were determined by fitting a model including these effects. That is why it is really important to take these effects into account. This acceleration is given by \cite{beu2005} :
\begin{equation}
\mathbf{A}_{GR}= \frac{k^2}{c^2r^3}\left[\left(4\frac{k^2}{r^3}-\left(\frac{d\mathbf{r}}{dt}\right)^2\right)\mathbf{r}+4\left(\mathbf{r}.\frac{d\mathbf{r}}{dt}\right)\frac{d\mathbf{r}}{dt}\right]
\end{equation}
where $\mathbf{r}$ is the heliocentric position vector of the comet, $k$ is the Gauss gravitational constant and $c$ the velocity of light.
I formulate the equation of motion in an equatorial heliocentric frame where  the variables are the cartesian coordinates of the position of the comet. The equations of the motion of the comet can be written as :
\begin{equation}
\frac{d^2\mathbf{r}}{dt^2}=-k^2\frac{\mathbf{r}}{r^3}+\sum_{i=1}^{8} k^2 m_i\left(\frac{\mathbf{r_i}-\mathbf{r}}{\|\mathbf{r_i}-\mathbf{r}\|^3}-\frac{\mathbf{r_i}}{\|\mathbf{r_i}\|^3}\right) + \mathbf{A}_{GR} + \mathbf{A}_{NG} \label{motion}
\end{equation}
where $m_i$ are the planet masses and $\mathbf{r_i}$ the heliocentric position vector of the $i^{th}$ planet.\\

Equation \ref{motion} can be reduced to a system of six first order differential equations as follow :
\begin{equation}
\frac{d\mathbf{X}}{dt}=f(\mathbf{X})\label{sys_motion}
\end{equation}
where the components $X_1$, $X_2$ and $X_3$ represent the components $x$, $y$ and $z$ of the heliocentric position vector $\mathbf{r}$ of the comet. $X_4$, $X_5$ and $X_6$ represent the components $v_x$, $v_y$ and $v_z$ of the velocity vector $\frac{d\mathbf{r}}{dt}$ of the comet.
The integrations of this system are performed using the 15-th order RADAU integator (\cite{eve1974}).

\subsection{Monte-Carlo method to clone the comet}
 By fitting the astrometrical measurements of the comet, it is possible to obtained the orbital elements. According to the least square method and due to the randomly gaussian errors of the observations, it is possible to determine the errors (see Table \ref{tab:ci}) of the orbital elements (or the cartesian coordinates by changing reference frame) and the non-gravitational parameters. \\
 
\begin{figure}
\begin{center}
\includegraphics[width=0.5\textwidth]{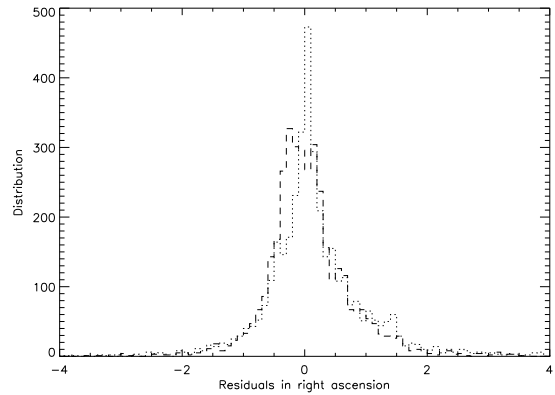}
\caption{Distribution of the optimal residual in right ascension (dotted line) and declination (dashed line)}\label{res}
\end{center}
\end{figure}

The chaotic behaviour of the comet is computed by considering several orbits with small variations of their initial conditions. The small variations are choosen randomly as :
\begin{equation}
X_i=X_{i0}+\epsilon_i \sigma_{X_{i0}}
\end{equation}
where $X_i$ are the initial conditions of the clone of the comet, $X_{i0}$ are the optimal initial conditions of the comet determined by fitting the astrometrical measurements, $\sigma_{X_{i0}}$ are the errors on $X_{i0}$ and $\epsilon_i$ is a random number choosen in the normal distribution of mean 0 and variance 1 (because of the presumed gaussian nature of the errors). I take care that the choosen values correctly describe the observations by rejecting the set of initial conditions with a too large \textit{rms} (typically larger than 2.5" for the orbit considering NGF and 3" without NGF. Above this limit, lots of observations are not described accurately by the dynamical model). As a comparison, the distributions of the residuals corresponding to the nominal solution are represented in Figure \ref{res}. The mean value of this residual is 0.11" and the \textit{rms} is 0.9". The initial conditions and their errors are presented in table \ref{tab:ci}. They come from the IMCCE database \footnote{\url{http://www.imcce.fr/fr/ephemerides/donnees/comets/FICH/CIF0029.php}}. These initial conditions are adjusted to 3730 observations from July 3, 1995 to November 15, 2014.\\

\begin{table}
 \centering
 \caption{Optimal initial conditions deduced by fitting the astrometrical measurements of 67P/C-G. These initial conditions come from the database of IMCCE (Notes on comets). The epoch of this elements is August 13, 2015 at 0h, julian date: 2457247.5} 
 \label{tab:ci}
 \begin{tabular}{cc}
 \hline
 Initial condition & Value and Uncertainty (1 $\sigma$) \\
 \hline
 $x$ [AU] & $0.5688522425\pm9.25\times10^{-6}$ \\
 $y$ [AU]& $1.0004591393\pm1.10\times10^{-5}$ \\
 $z$ [AU]& $0.4703160954\pm4.98\times10^{-6}$\\
 $v_x$ [AU.$\mathrm{day}^{-1}$]& $-0.0174785349\pm1.16\times10^{-7}$\\
 $v_y$ [AU.$\mathrm{day}^{-1}$]& $0.0072390883\pm9.04\times10^{-8}$\\
 $v_z$ [AU.$\mathrm{day}^{-1}$]& $0.0057134963\pm5.03\times10^{-8}$\\
 $A_1$ & $0.10875\times10^{-8}\pm0,00092\times10^{-8}$\\
 $A_2$ & $0.01092\times10^{-8}\pm0,00002\times10^{-8}$\\
 $A_3$ & -\\
 \hline
 \hline
 Time of perhelion & August, 13.08694 2015 TT $\pm0.00000$ TT \\
 a [AU] & $3.4619083\pm0.0000161$\\
 e & $0.6408730\pm0.0000015$\\
 i [deg]  & $7.0402569\pm0.0000161$ \\
 $\omega$ [deg] & $12.7961939\pm0.0001227$ \\
 $\Omega$ [deg] & $50.1354102\pm0.0001417$ \\
 n [deg.$\mathrm{day}^{-1}$] & $0.15301385\pm0.00000107$ \\
 $\mathrm{q_h}$ [AU] & $1.2432648\pm0.0000005$ \\
 \hline
 \end{tabular}
 \end{table}
 
\section{Results}\label{results}

\begin{figure*}
{\includegraphics[width=0.48\textwidth]{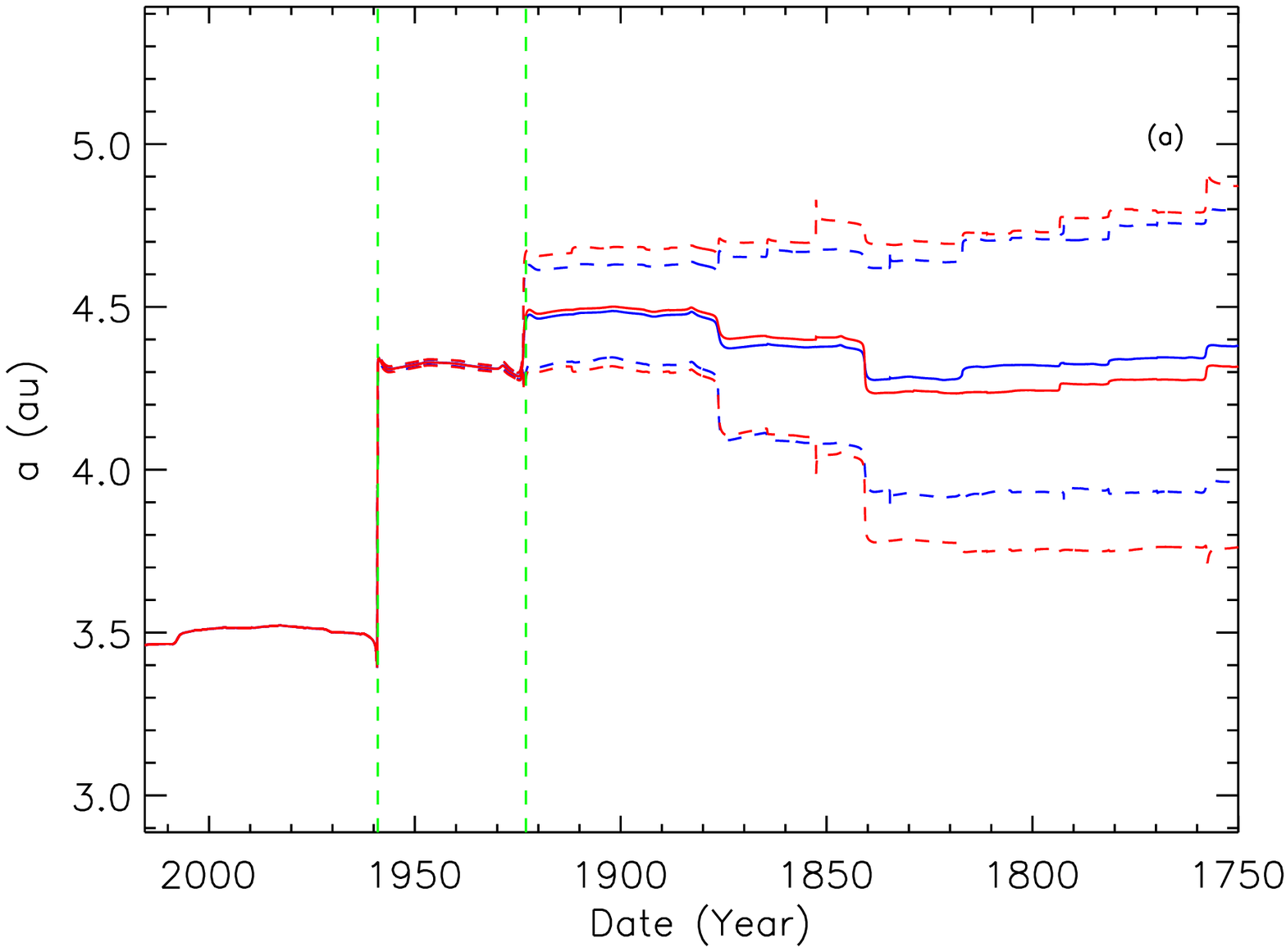}\phantomsubcaption\label{a_s}}
{\includegraphics[width=0.48\textwidth]{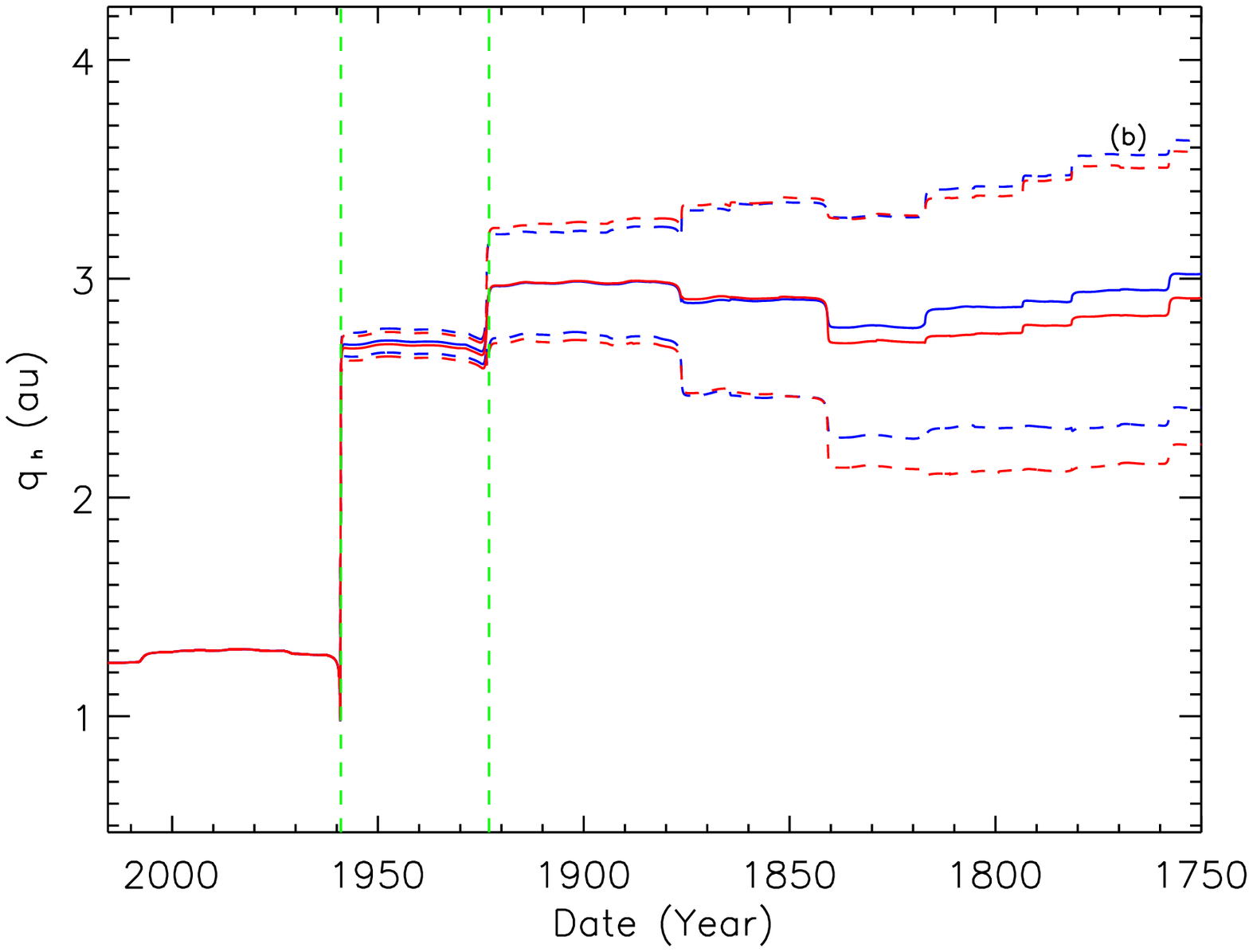}\phantomsubcaption\label{q_s}}
{\includegraphics[width=0.48\textwidth]{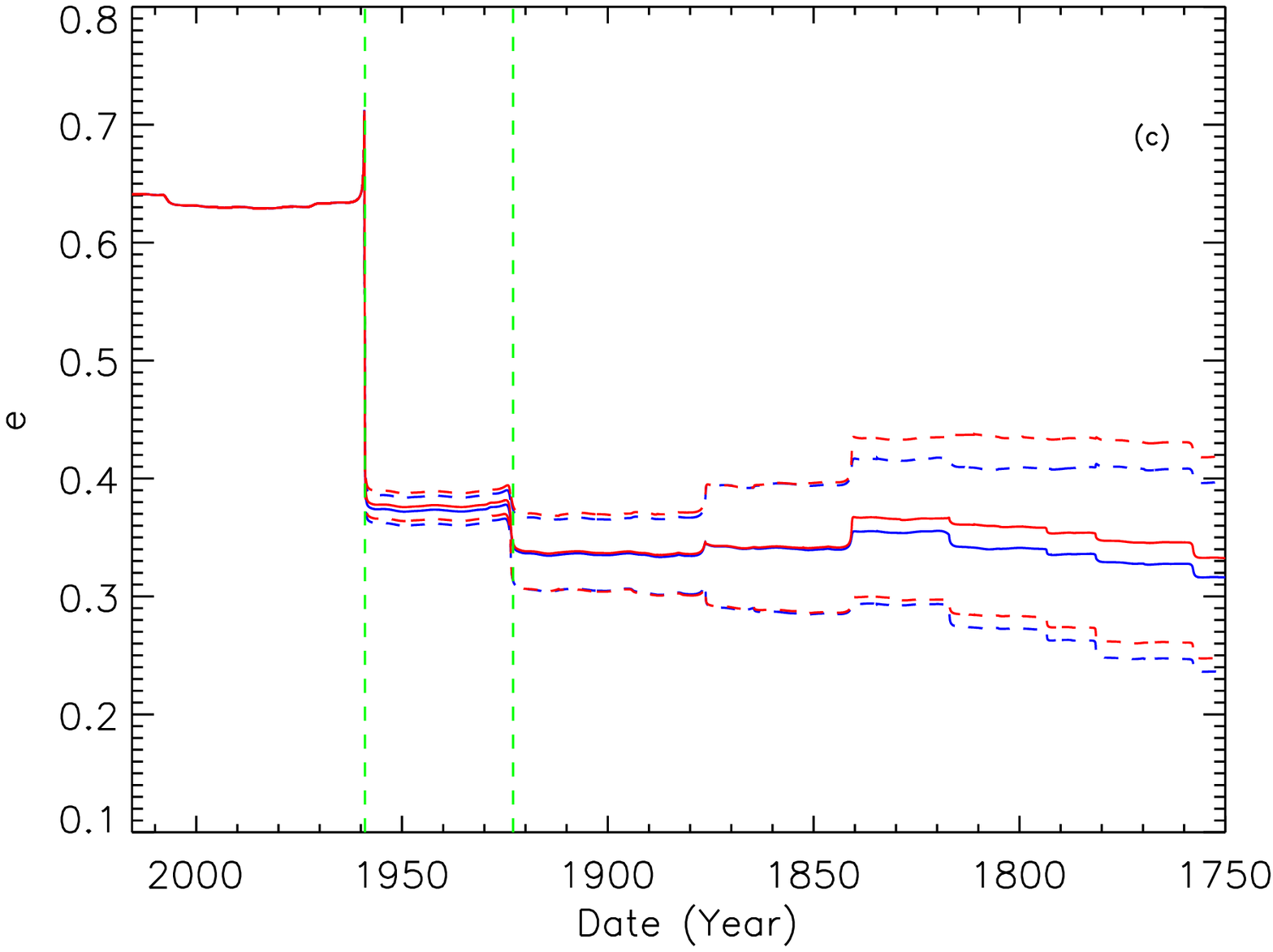}\phantomsubcaption\label{e_s}}
{\includegraphics[width=0.48\textwidth]{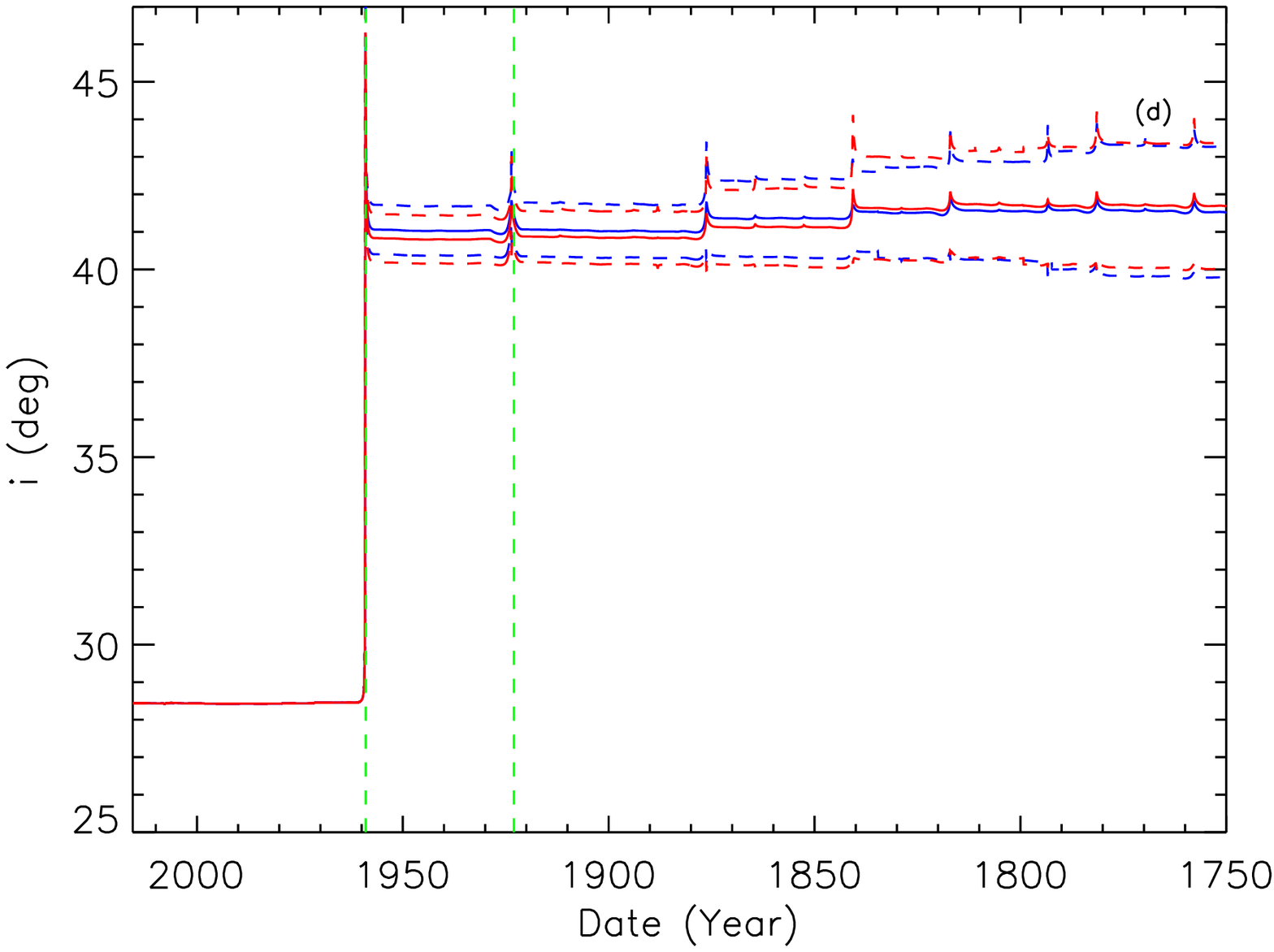}\phantomsubcaption\label{i_s}}
{\includegraphics[width=0.48\textwidth]{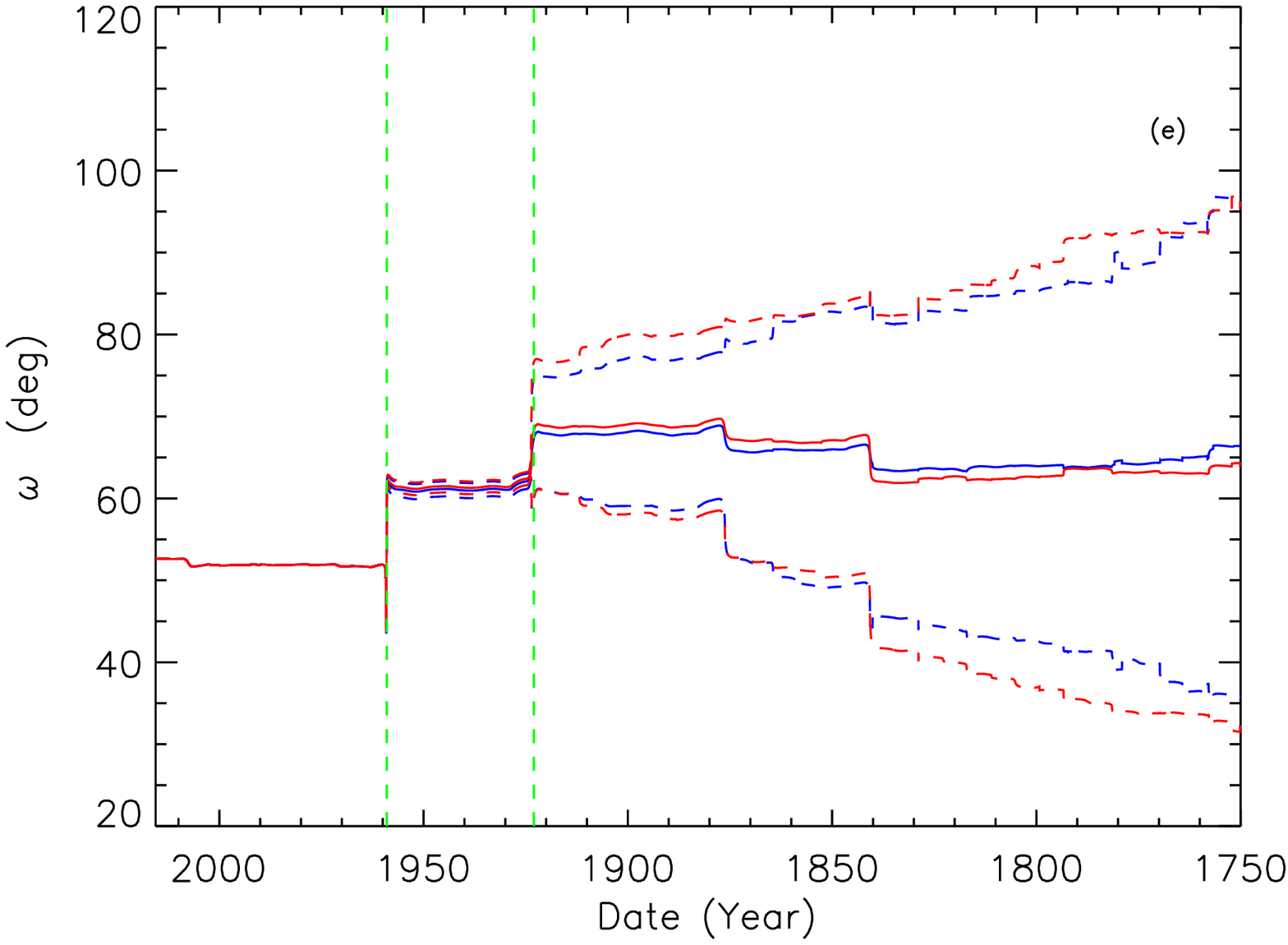}\phantomsubcaption\label{om_s}}
{\includegraphics[width=0.48\textwidth]{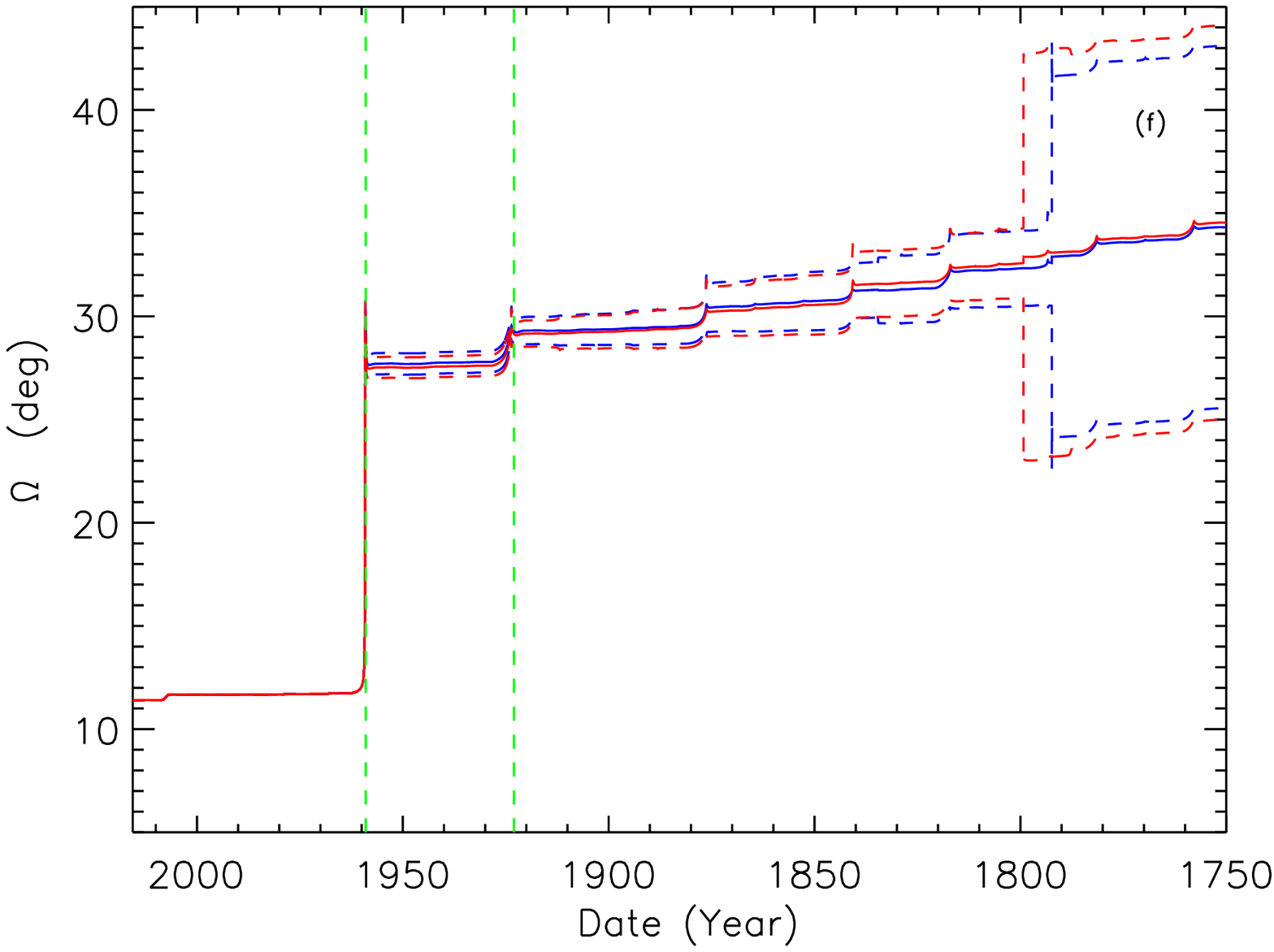}\phantomsubcaption\label{gom_s}}
\caption{Variation of the orbital elements ($a$, $q$, $e$, $i$, $\omega$ and $\Omega$) over 275 years. Plain line is the mean value of the element over 1000 orbits and dashed lines represent the standard deviation. In blue are the orbits with NGF and in red the orbits without NGF. The green dashed lines represent the two last close encounter with Jupiter.}\label{orb_short}
\end{figure*}

\begin{figure*}
{\includegraphics[width=0.48\textwidth]{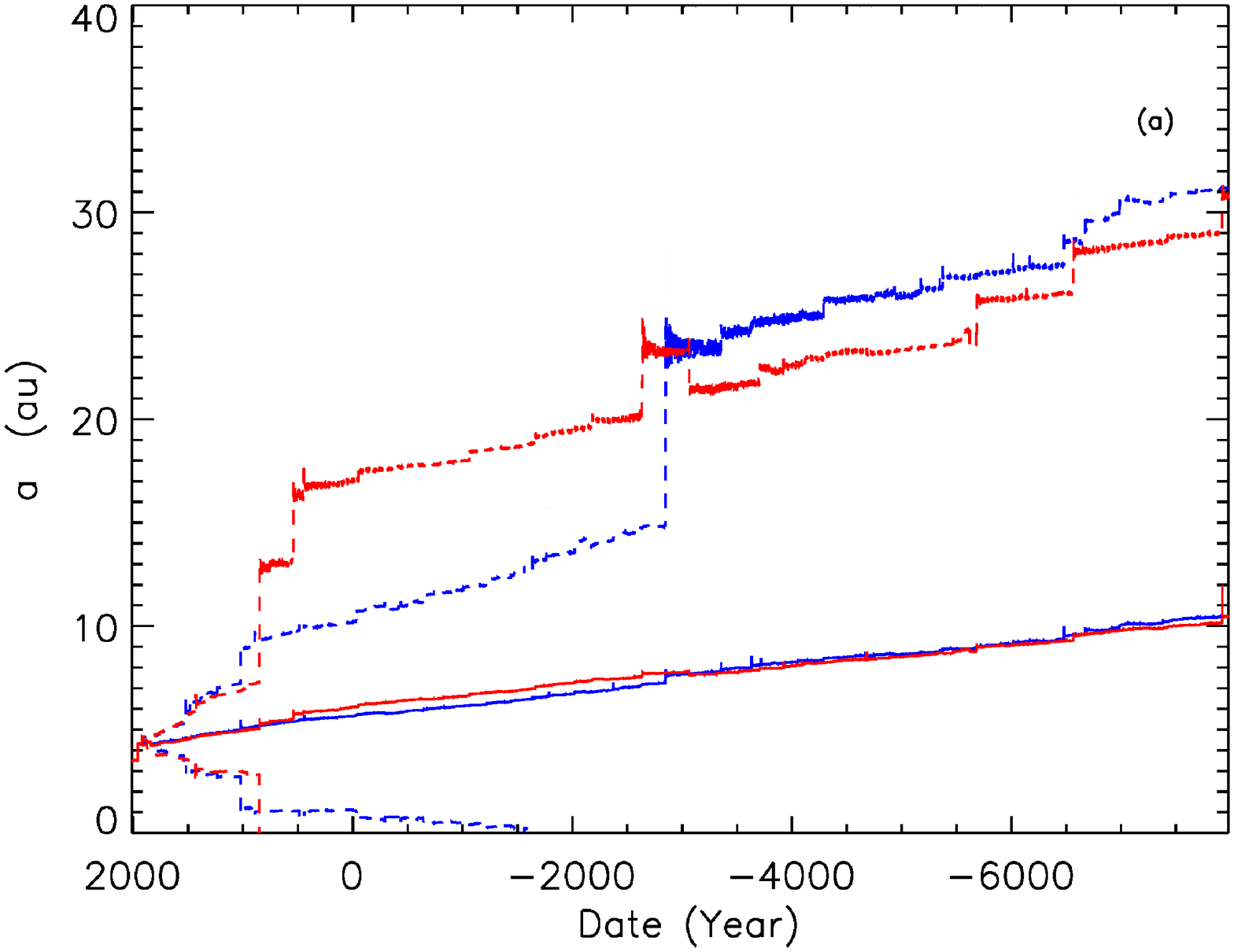}\phantomsubcaption\label{a_ls}}
{\includegraphics[width=0.48\textwidth]{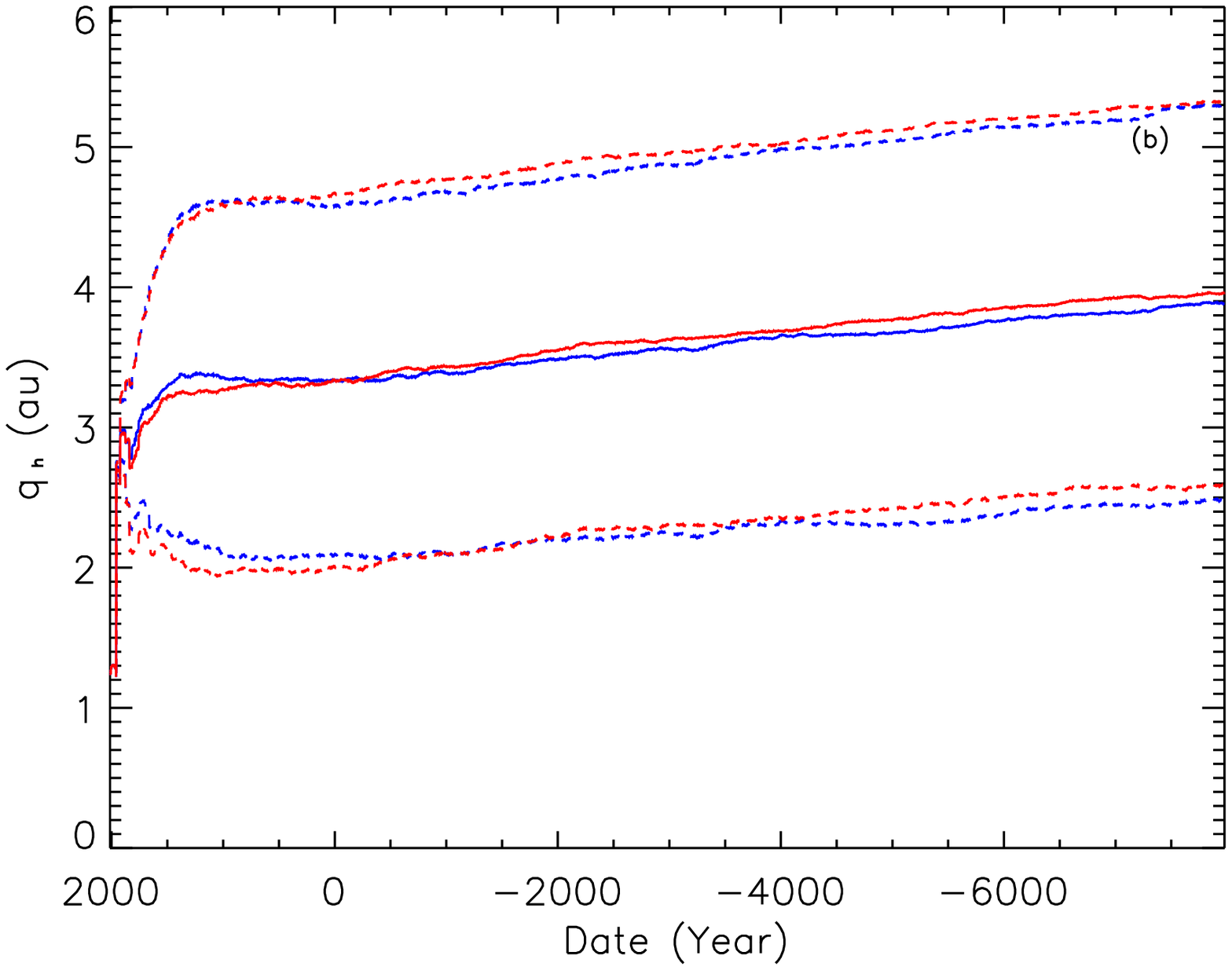}\phantomsubcaption\label{q_ls}}
{\includegraphics[width=0.48\textwidth]{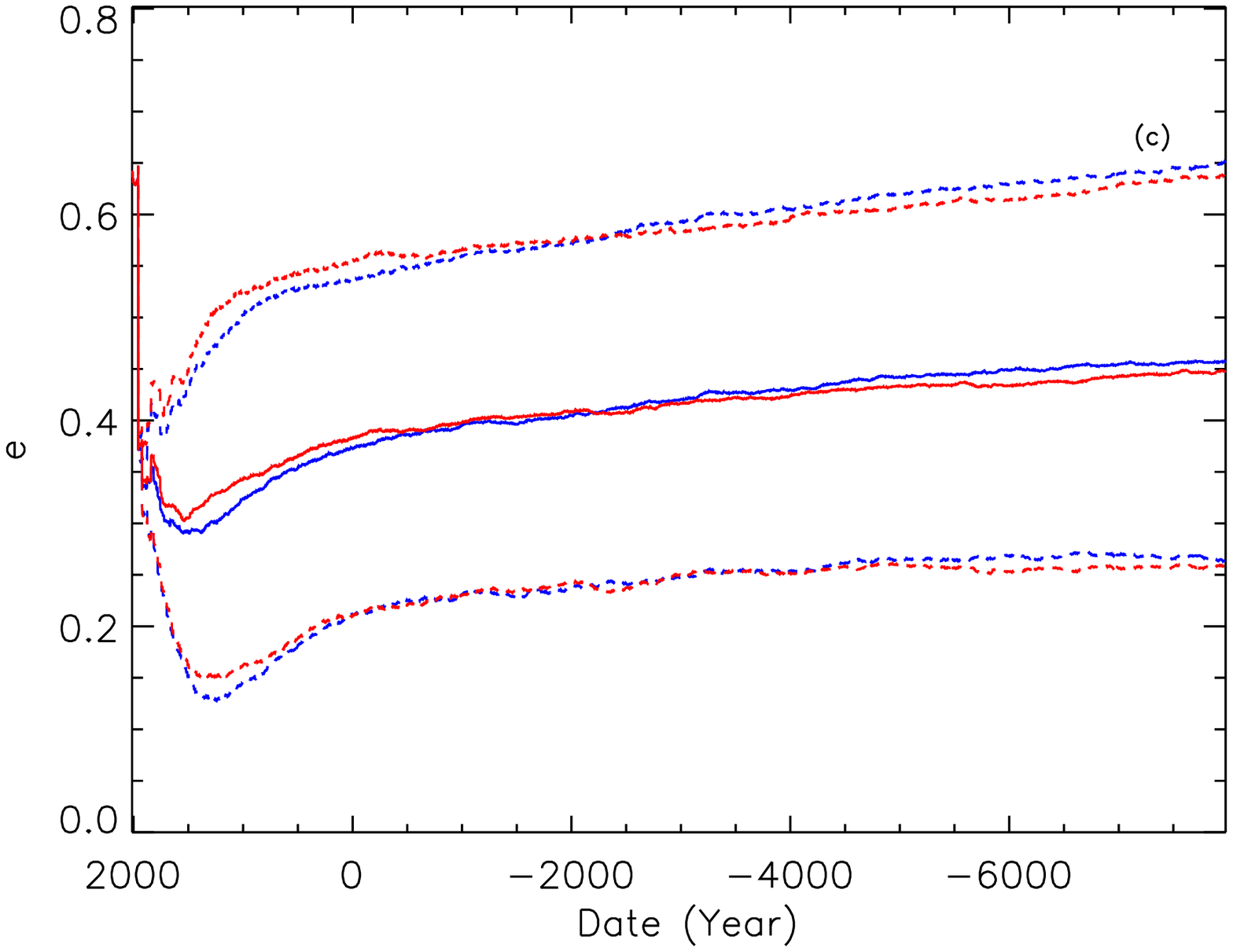}\phantomsubcaption\label{e_ls}}
{\includegraphics[width=0.48\textwidth]{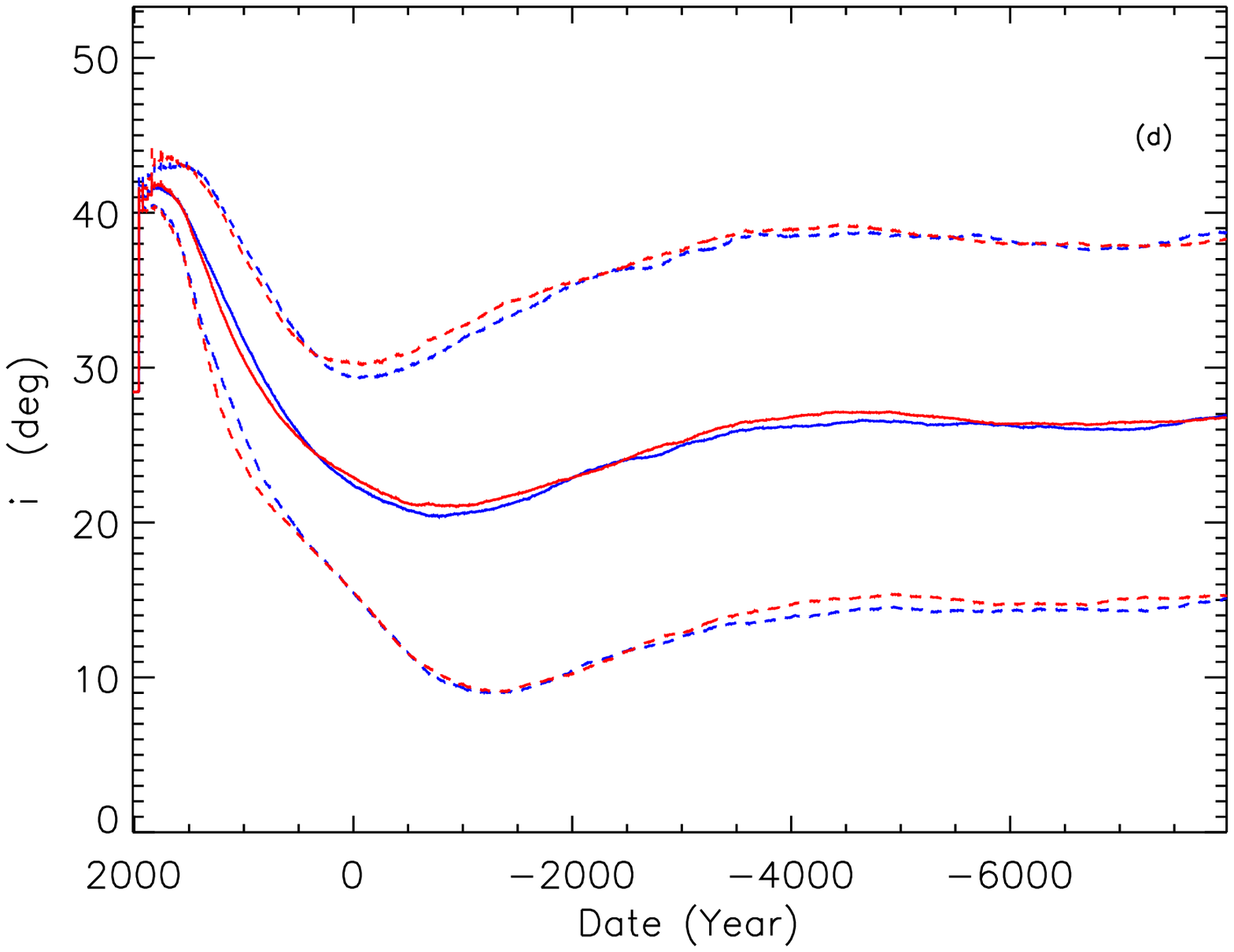}\phantomsubcaption\label{i_ls}}
{\includegraphics[width=0.48\textwidth]{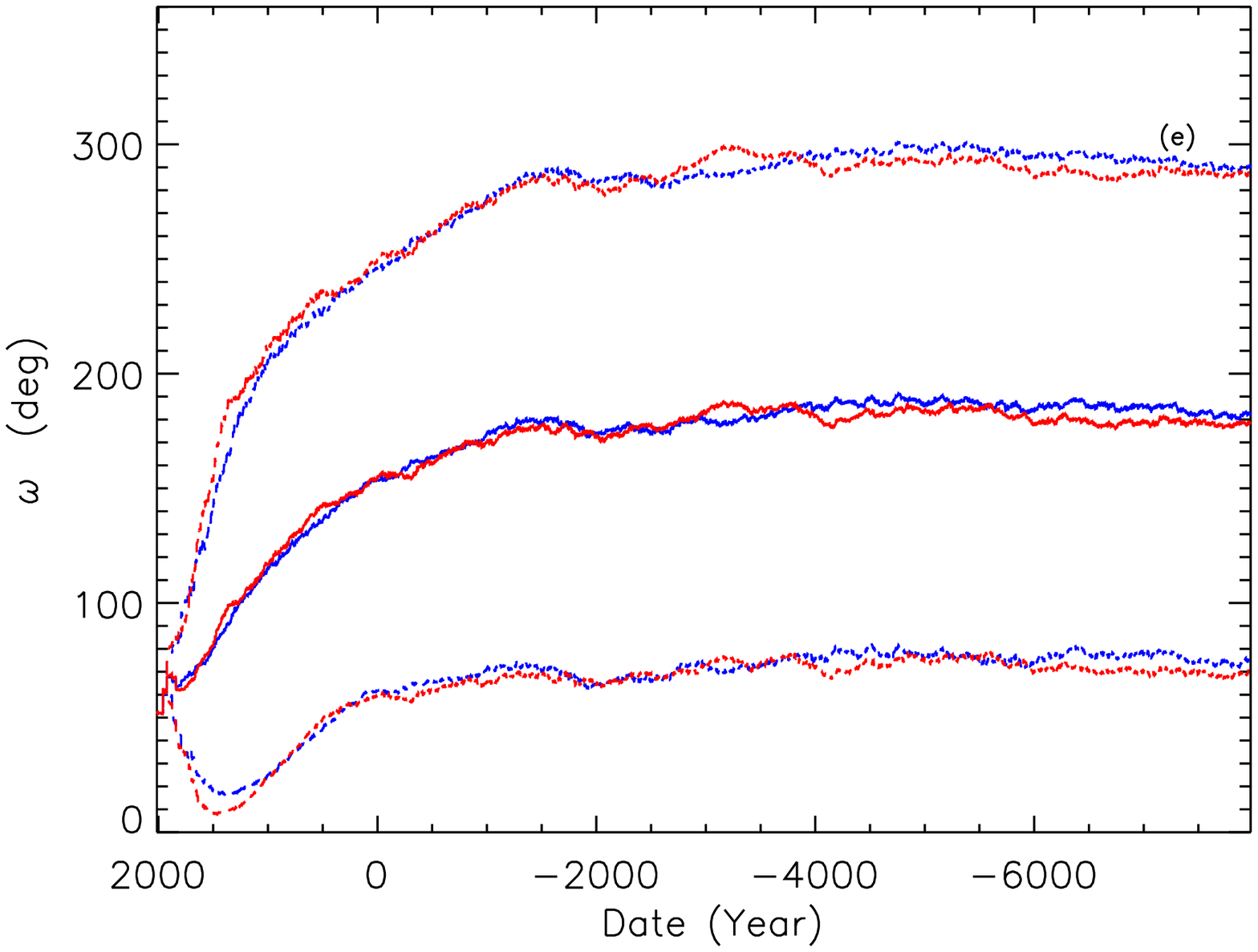}\phantomsubcaption\label{om_ls}}
{\includegraphics[width=0.48\textwidth]{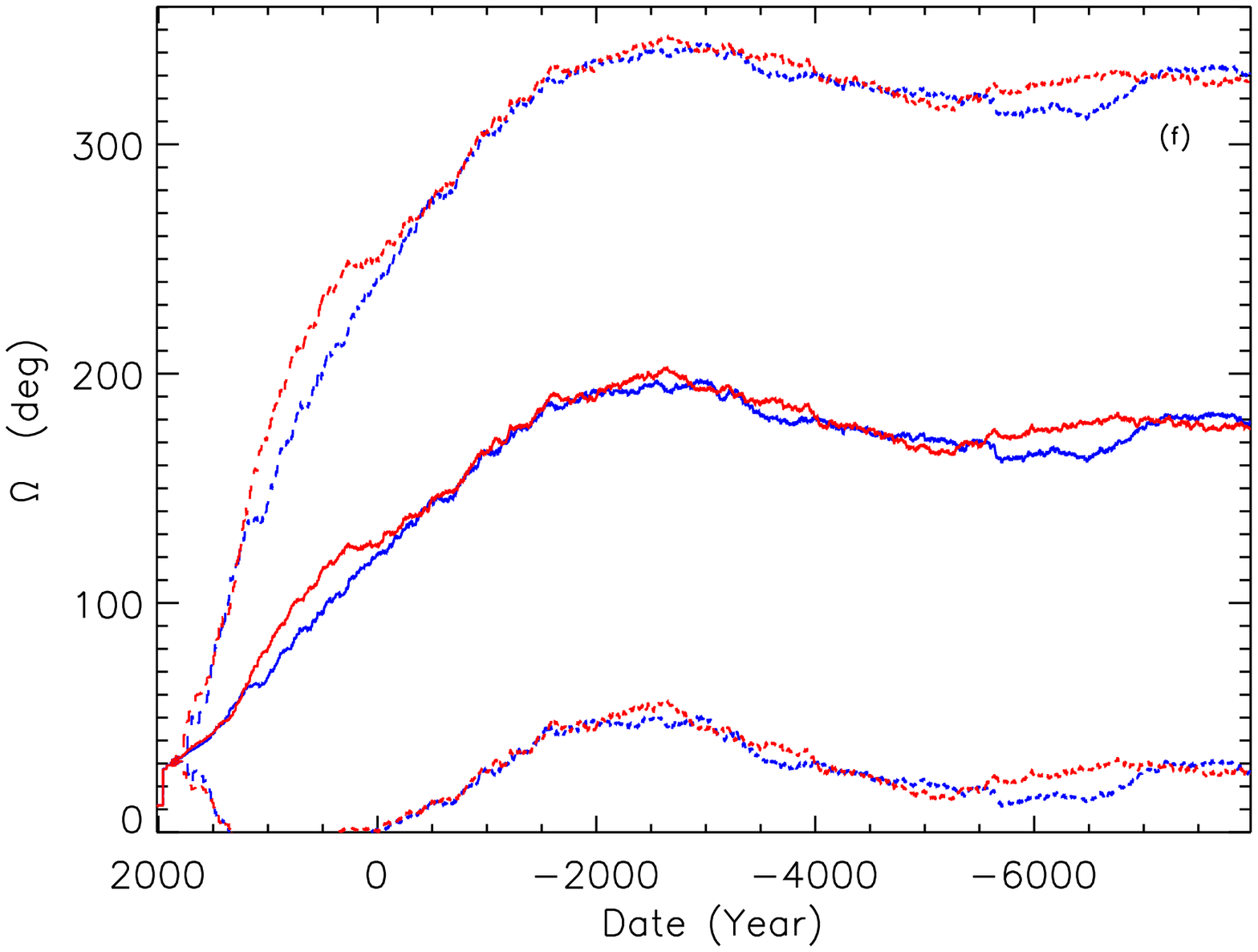}\phantomsubcaption\label{gom_ls}}
\caption{Variation of the orbital elements ($a$, $q$, $e$, $i$, $\omega$ and $\Omega$) over 10,000 years. Plain line is the mean value of the element over 1000 orbits and dashed lines represent the standard deviation. In blue are the orbits with NGF and in red the orbits without NGF. }\label{orb_long}
\end{figure*}

The orbits are backward integrated firstly with 10000 steps of 10 days which is almost 275 years and secondly with 10000 steps of 1 year which correspond to 10000 yr. For each time period, 2000 clones were integrated (1000 with non-gravitational forces and 1000 without). For each set of 1000 orbits, the mean value (plain lines) and the standard deviation (dashed lines) of the six orbital elements are computed  and presented on figures \ref{orb_short} and \ref{orb_long} . According to the definition of the normal law, we note that 68$\%$ (which correspond to the statistic to be between the mean plus or minus the standard deviation) of the orbits lie between the dashed lines of the figures. The considered six orbital elements are : the semi major axis $a$, the eccentricity $e$, the orbit inclination $i$, the argument of the perihelion $\omega$, the longitude of the ascending node $\Omega$ and the perihelion distance $q_h$. \\

\begin{figure}
\begin{center}
\includegraphics[width=0.5\textwidth]{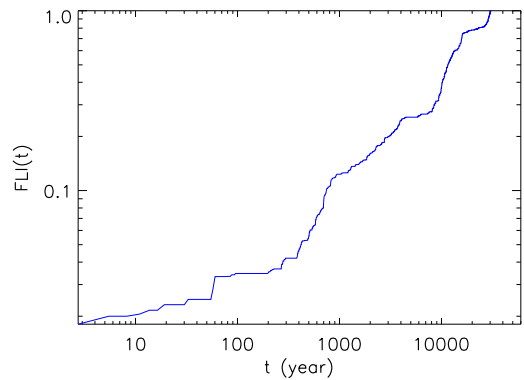}
\caption{Evolution the Fast Lyapunov indicator as a function of the time. }\label{lyap}
\end{center}
\end{figure}

The comet is known to experience close encounter with Jupiter. The two last closest encounters happen on February $\mathrm{4^{th}}$, 1959 (closest distance 0.05 AU) and October $\mathrm{2^{nd}}$, 1923 (closest distance 0.92 AU).
The 1959 close approach drastically modified the orbit of the comet. The semi-major axis was reduced from about 4.3 AU to 3.5 AU and its perihelion distance was reduced from about 2.7 AU to 1.3 AU (see Fig. \ref{a_s} and \ref{q_s}).The semi-major axis and the perihelion distance were not the only parameter affected by by this event. Indeed, the orbit became more eccentric ($e$ increased form 0.38 to 0.63) and closer to the ecliptic plane ($i$ decreased from $41^{\circ}$ to $28^{\circ}$). 
Every close encounter of the comet with Jupiter induce less precision of the past comet orbits. Indeed, each close approach tends to increase the chaoticity of the orbit. The chaoticity of the motion of the comet was investigated using the Fast Lyapunov Indicator (here after FLI). The FLI is defined as (\cite{fou2002,fro1997}):

\begin{equation}
\text{FLI(t)}=\sup_{0<k<t} \ln \|\mathbf{w}(t)\| \label{FLI}
\end{equation}

where the vector $\mathbf{w}(t)$ is the solution of the variational equation $
\frac{d\mathbf{w}}{dt}=\frac{\partial f(\mathbf{X})}{\partial\mathbf{X}}\mathbf{w}$ of the system (Equation \ref{sys_motion}). Figure \ref{lyap} shows the evolution of the FLI as a function of the time. As the FLI increases exponentially in a log-log scale, we can conclude that the system is chaotic (\cite{fou2002,fro1997}).

For a date anterior to the close encounter with Jupiter in 1923, the two sets (with and without NGF) of 1000 orbits are diverging relatively to each other. Indeed, before 1923, the orbital element can take values in a large range (see Fig. \ref{orb_short} and \ref{orb_long}). This is due to the fact that for a time anterior to the date of the 1923 close encouter, the system was chaotic. \\

Nevertheless, it is possible to study the trends of the orbital elements in the past. We note that, in the past, the orbit was surely a bit more circular (see Fig. \ref{e_s} and \ref{e_ls}) than now with an eccentricity around 0.4. We can note a large variation of the inclination during the period $[-1000;1960]$ years. The inclination increased from about $25^{\circ}$ to about $42^{\circ}$ (see Fig. \ref{i_ls}). \\
According to figure \ref{a_ls} and \ref{q_ls}, we see that the comet was probably orbiting under the control of Jupiter as noted by \cite{kro2003} with a perihelion distance of about 3.5 au and a semi-major axis varying from 10 au 10,000 years backward to 4.5 au prior the close encouter with Jupiter in 1923. We can note that the NGF does not have a big influence on the evolution of the orbital elements of the comet except for the dispersion of the semi-major axis (see Fig. \ref{a_ls}). \\
According to figure \ref{om_s}, \ref{gom_s}, \ref{om_ls} and \ref{gom_ls}, the plane of the orbit and the orbit itself are probably slowly precessing.\\

Finally, according to figure \ref{q_s} and \ref{q_ls}, we see that before the close encounter with Jupiter in 1959, the perihelion distance was further from the sun than it is today. Indeed, at least 84\% of the calculated orbits (orbit beyond $(\bar{q_h}-\sigma_{q_h})$) are lying beyond 2 AU where the activity of the comet is low (see fig. \ref{g_r}). So the nucleus surface may be young and not much altered by outgassing processes.

\section{Conclusion}\label{conclu}
This study shows the mean evolution and the dispersion of the different orbital elements of the comet 67P/Churyumov-Gerasimenko, target of the ESA/Rosetta mission. This is done for two sets of 1000 clone orbits considering or not the non-gravitational forces for two time scale : 275 years and 10,000 years. This work is usefull to interpert the observations of the Rosetta spaceprobe.
Indeed, the main conclusions of this study are the followings :
\begin{itemize}
\item The close encounter with Jupiter on February $\mathrm{4^{th}}$, 1959 drastically modified the orbital elements of the comet.
\item The motion of the comet is chaotic in the past before the 1923 close encounter with Jupiter.
\item According to the mean trend of $q_h$ and for a date anterior to the 1959 close encounter with Jupiter, the comet was orbiting farther from the sun ($q_h>2$ AU) where the nucleus activity (outgassing) was lower than now. Thus, the material at the nucleus surface has to be relatively young.
\item The variation of the mean orbital element are not really sensitive to the effects of the non-gravitational forces except for the distribution of the clone orbit for the semi-major axis.
\end{itemize}

\section*{Acknowledgments}
It is a great pleasure to thank Marc Fouchard and the anonymous referees for very useful comments and advice. I thank also Jérémie Vaubaillon and Frédéric Pierret for their carefull proofreading. Finally, I want to thank Jacques Laskar, Philippe Robutel and Mickael Gastinau for their advice concerning the chaos indicator and for providing me a long term solution of the planetary theory INPOP.\\

I am supported by a Europeen Space Agency (ESA) research fellowship at the Europeen Space Astronomy Center (ESAC), in Madrid, Spain.

\bibliographystyle{plain}
\bibliography{chaos_67P_bib}

\end{document}